\documentclass[conference]{IEEEtran}

\usepackage{amsmath,amssymb,epsfig,cite,footnote,authblk}

\newcommand{\pr}[1]{\left( #1\right)}
\newcommand{\prr}[1]{\left[ #1 \right]}
\newcommand{\es}[1]{\begin{equation}\begin{split}#1\end{split}\end{equation}}

\newcommand{\R}{\mathbb{R}}

\newcommand{\C}{\mathbb{C}}

\newcommand{\V}{\mathcal{V}}

\newcommand{\PP}{\mathcal{P}}

\newcommand{\br}{\mathbf{r}}
\newcommand{\bt}{\mathbf{t}}
\newcommand{\dd}{\textrm{d}}

\DeclareMathOperator*{\argmax}{arg\,max}

\begin{document}

\title{Distributed Power Allocation and Channel Access Probability Assignment for Cognitive Radio}
\author[]{Orestis Georgiou}
\author[]{Mohammud Z. Bocus}
\author[]{Shanshan Wang}
\affil[]{Toshiba Telecommunications Research Laboratory, 32 Queens Square, Bristol, BS1 4ND, UK.}
\maketitle


\begin{abstract}
In this paper, we present a framework for distributively optimizing the transmission strategies of secondary users in an ad hoc cognitive radio network.  In particular, the proposed approach allows secondary users to set their transmit powers and channel access probabilities such that, on average, the quality of service of both the primary and secondary networks are satisfied.  The system under consideration assumes several primary and secondary transceiver pairs and assumes no cooperation or information exchange between neither primary and secondary users nor among secondary users.  The outage probability, and consequently the connection probability, is derived for the system and is used in defining a new performance metric in the optimization problem using tools from stochastic geometry.  We refer to this metric as the spatial density of successful transmission.  We corroborate our derivations through numerical evaluations.  We further demonstrate that even in the absence of any form of cooperation, an acceptable quality of service can be attained in the cognitive radio environment.
\end{abstract}

\section{Introduction \label{sec:intro}}
To address the issues of scarce spectrum resources and satisfy the ever-increasing demand for higher data rate wireless services, cognitive radio technologies are being considered for future communication networks.  While there is a physical limitation on the amount of spectrum that is available at any instant, it is realized that the main problem with the vast majority of communication systems today is the inefficient usage of the allocated spectral resources.  In fact, studies by regularization bodies such as the FCC has shown that a large portion of the bandwidth assigned remain under utilized while unallocated spectrum is limited.
Cognitive radio \cite{Mitola1999a} is the paradigm shift in wireless communications wherein secondary, unlicensed users are allowed to communicate over the bandwidth licensed to the primary users provided that the quality of service (QoS) degradation of the primary users are within acceptable bounds.

Since the concepts of cognitive radio and dynamic spectrum access were introduced to the communications community, their promised gains and means to achieve them have been extensively studied in the literature (see e.g., \cite{Cumanan2010, DelRe2009, Fu2008} and references therein).  The investigated research topics on cognitive radio include, but are not limited to, power allocation for secondary users, interference cancellation and alignment, multiuser scheduling, multiple antenna systems, among others.  The aim of this paper is to provide further insights in the design and analysis of cognitive radio systems.  In particular, power allocation strategies and random channel access schemes are presented under a stochastic geometry framework \cite{haenggi2009stochastic}; the study of random spatial patterns.

Power allocation and channel access protocols for secondary users are important design parameters for maintaining the QoS requirements of the licensed users.  While centralized power control techniques can lead to the optimal performance, achieving such gains in practice may not be possible.  For instance, having the central server or device collect all the channel state information from both primary and secondary users entails large signalling overhead, delays and processing complexities.  Optimization problems in such scenarios are often non-convex in nature and do not lend themselves to efficient, low-complexity solutions.  Moreover, in practical scenarios, it is unlikely that the primary system would continuously exchange channel state information with the cognitive users.  On that account, several works have considered the possibility of decentralized or distributed power allocation \cite{Wu2008,Yan2012,Wu2009,Seneviratne2011,Ahmed2014}.  The strategy for solving the problem of distributed resource allocation has generally relied on game theoretic approaches.  In \cite{Wu2008}, the distributed multichannel power allocation problem for cognitive radio users is formulated as a non-cooperative game.  Similarly, the work in \cite{Seneviratne2011} uses a game theoretic approach for resource allocation in cognitive radio networks with non-cooperative spectrum sharing.  On the other hand, in \cite{Nadkar2012b}, the authors present a distributed power allocation strategy for cognitive radio users  where secondary users are allowed to exchange information and negotiate the best operating point. The latter work, in contrast to most others in the literature, consider a spectrum overlay cognitive radio system.

In this paper, we do not follow a game theoretic approach for solving the problem of distributed power allocation in cognitive radio systems.  Rather, the concept of network connectivity \cite{georgiou2014network,keyhole} and interference effects as captured using tools from stochastic geometry \cite{haenggi2009stochastic} is used to quantify the performance of dense ad hoc networks.  
Namely, the outage probability, and consequently the connectivity, of the cognitive radio system in the presence of interference from both primary and secondary users is derived.  
The derived closed form expressions, along with the constraints imposed by the primary users, are then used to specify the power allocation strategy at each secondary transmitter.  
In addition, we consider the effects of random channel access in the connectivity analysis and derive the optimal channel access probability of the secondary users under given constraints.  
Optimization is performed on both these variables and a simple process is proposed that can run locally on each secondary transmitter.
This work differs from existing literature in that we analyze and exploit the random location of nodes (transmitters and receivers) along with other randomness in the channels (e.g. fading and channel access) to arrive at the most adequate operating point.  

The rest of the paper is structured as follows.  In Section \ref{sec:model}, the system model, definitions and assumptions are presented.  The outage analysis of the network under consideration is given in Section \ref{sec:outage}.  The distributed optimization framework is given in Section \ref{sec:CR}, while some simulation results and evaluations are provided in Section \ref{sec:sim}.  Finally, some concluding remarks are given in Section \ref{sec:con}. 

\section{Network Definitions and System Model \label{sec:model}}

\subsection{Primary and Secondary transmitter and receiver pairs}
Consider a cognitive radio ad hoc network composed of primary and secondary users where the latter can only use the licensed spectrum when transmissions do not degrade the QoS requirements of the primary network.
The primary transmitters (PTs) are uniformly distributed in space and therefore we model their locations by a homogeneous Poisson point process (PPP) in $\R^2$ with density $\lambda_{\alpha}$.
We denote the location of primary transmitter $i$ by $\bt_i^\alpha \in \R^2$ and its transmit power by $\mathcal{P}_i^\alpha$ for $i=1,2,\cdots,N_\alpha$, where $N_\alpha$ is a Poisson random variable.
Each PT is associated with a primary receiver (PR) which is uniformly distributed within a given range $R$.
It follows that the spatial distribution of the PRs also follows a two-dimensional PPP with density $\lambda_{\alpha}$.
We denote the location of the PR associated to PT $i$ by $\br_{i}^\alpha \in \R^2$ for $i=1,2,\cdots,N_\alpha$.

\begin{figure}[t]
\centering
\includegraphics[width=0.45\textwidth]{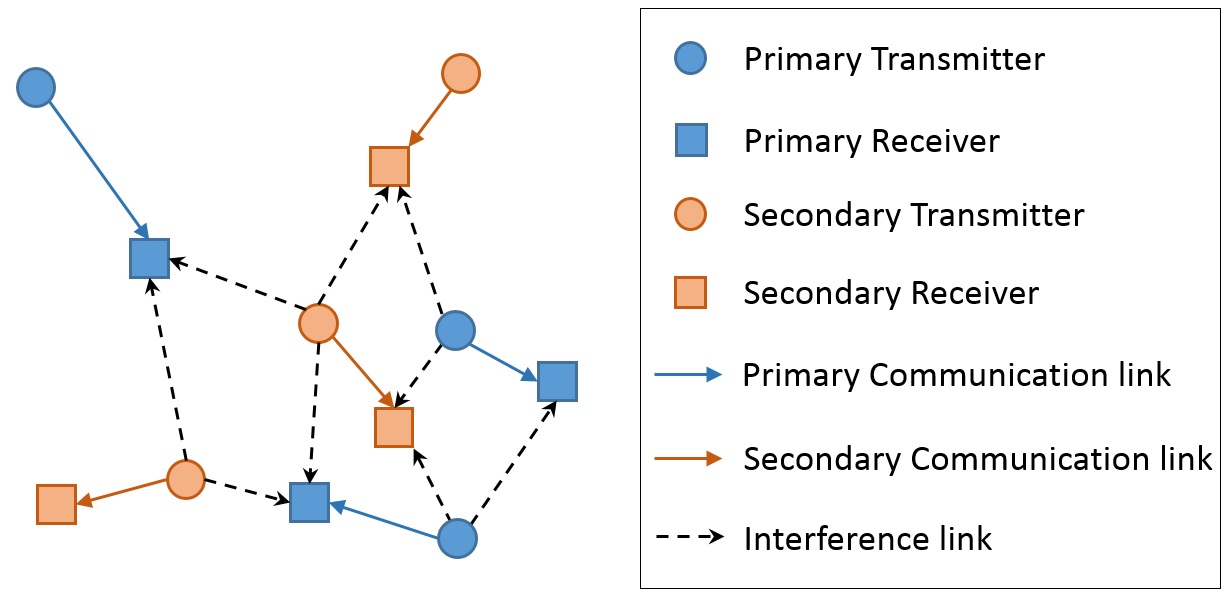}
\caption{
Network model showing three primary and three secondary transmitter-receiver pairs (12 nodes in total).
Interference links are also shown between primary (secondary) transmitters and secondary (primary) receivers as well as primary (secondary) unassociated receivers.
Not all interference links are shown for the sake of clarity.
}
\label{fig:model}
\end{figure}

A similar setup is used to model the secondary transmitters (STs) and receivers (SRs) with density $\lambda_\beta$. 
We denote the location of secondary transmitter $i$ by $\bt_i^\beta \in \R^2$, its associated receiver by $\br_{i}^\beta \in \R^2$ and its transmit power $\mathcal{P}_i^\beta$ for $i=1,2,\cdots,N_\beta$. 
Finally, the distance between a transmitter and a receiver is given by $d_{ij}^{xy}=| \bt_i^x - \br_j^y |$ for $x,y = \{\alpha,\beta\}$.
Note that superscripts (subscripts) of $\alpha$'s and $\beta$'s ($i$'s and $j$'s) are used throughout the paper and do not indicate raising to a power but simply distinguish between primary and secondary users (transmitters and receivers) respectively.
The subscript and superscript order in $d_{ij}^{xy}$ is very important as it distinguishes between transmitter / receiver, and network tier. 
For example, $d_{12}^{\alpha\beta}$ is the Euclidean distance separating the primary transmitter $1$ located at $\bt_1^\alpha$ and the secondary receiver $2$ located at $\br_2^\beta$. 
A schematic of this setup is shown in Fig. \ref{fig:model}.

\subsection{Path-loss Attenuation and Fading}
Connectivity and capacity of dense ad hoc networks strongly depend on the behaviour of the underlying propagation medium and modelled through the attenuation function \cite{dousse2004connectivity}.
Here, the wireless channel attenuation is modelled as the product of a small-scale fading component and a large-scale path-loss component. 
The latter follows from the Friis transmission formula where the long time average signal-to-noise ratio (SNR) at a receiver (in the absence of interference) decays with distance like $\textrm{SNR}_{ij}^{xy}\propto (d_{ij}^{xy})^{-\eta}$, where $\eta$ is the path loss exponent usually taken to be $\eta=2$ in free space and $\eta>2$ in cluttered urban environments. 
We therefore define a path-loss function $g(d_{ij}^{xy})$ given by \cite{gong2014interference}
\es{
g(d_{ij}^{xy})= \frac{1}{\epsilon + (d_{ij}^{xy})^\eta} , \qquad \epsilon\geq0
\label{path}
.}
The $\epsilon$ buffer is included to make the path-loss model non-singular at $d_{ij}^{xy}=0$.
The small-scale fading component is assumed Rayleigh such that the channel gain $|h_{ij}^{xy}|^2$ between a transmitter $i$ and receiver $j$ is modelled by an exponential random variable of mean $1$.
Note that the channel transfer coefficient $h_{ij}^{xy}\in\C$ is assumed symmetric (i.e., $h_{ij}^{xy}=h_{ji}^{yx}$).

\subsection{Random Access Channel}
Device locations and wireless channel fading are two sources of randomness characteristic of ad hoc networks.
A third one is the channel access scheme employed.
ALOHA \cite{metcalfe1976ethernet} and CSMA \cite{kleinrock1975packet} are two well accepted classes of random and distributed MAC protocols.
In this paper, a protocol similar to the former is adopted, wherein each node transmits randomly, irrespective of any nearby transmitter \cite{baccelli2006aloha}.  This has the effect of independently thinning the spatial PPP node distribution.  On this note, we define the parameter $\chi_i^x= \{0,1\}$ for $x=\{\alpha,\beta\}$, a Bernoulli random variable with mean $p_i^x \in[0,1]$ indicating whether device $i$ transmits during a particular resource block or not.
For example, if $p_i^\beta = 1/3$, for all secondary transmitters, then approximately a third of the secondary transmitters will be active at any given time instance.

\subsection{Primary and Secondary SINR's}
We now turn to our main metric of interest, the signal-to-interference-plus noise-ratio ($\textrm{SINR}$) at a given receiver; a proxy to the average link throughput.
For the $i$th primary transmitter-receiver pair we have 
\es{
\textrm{SINR}_{i}^{\alpha} = \frac{\mathcal{P}_i^\alpha |h_{ii}^{\alpha\alpha}|^2 g(d_{ii}^{\alpha\alpha})}{\mathcal{N} +  \mathcal{I}_i^{\alpha\alpha} +  \mathcal{I}_i^{\beta\alpha}   }
\label{SINR1}
,}
where $\mathcal{N}$ is the average background noise power, and $\mathcal{I}_i^{\alpha\alpha}$ and $\mathcal{I}_i^{\beta\alpha}$ is the interference received at receiver $i$ from primary and secondary transmitters respectively such that
\es{
\mathcal{I}_i^{\alpha\alpha} &= \gamma_{\alpha\alpha} \sum_{k\not=i} \chi_k^\alpha \mathcal{P}_k^\alpha |h_{ki}^{\alpha\alpha}|^2 g(d_{ki}^{\alpha\alpha})\\
\mathcal{I}_i^{\beta\alpha} &= \gamma_{\beta\alpha} \sum_{k} \chi_k^\beta \mathcal{P}_k^\beta |h_{ki}^{\beta\alpha}|^2 g(d_{ki}^{\beta\alpha})
\label{inter}
.}
The SINR's for the $i$th secondary transmitter-receiver pair are obtained by simply exchanging $\alpha \leftrightarrow \beta$ in \eqref{SINR1} and \eqref{inter}.
The factors $\gamma_{xy}\in[0,1]$ in \eqref{inter} serve as weights for the interference terms and model the gain of the spread spectrum scheme used (if any). 
For instance, in a broadband CDMA scheme $\gamma$ depends on the code orthogonality with $\gamma_{xy}=1$ corresponding to a narrow band system \cite{yang2012connectivity}.
Alternatively, when $\gamma_{xy}=0$ the signal from node $\bt_i^x$ is completely orthogonal to all other tier $x$ concurrent transmissions.
In this paper we consider four types of interference captured by the factors $\gamma_{\alpha\alpha}$, $\gamma_{\alpha\beta}$, $\gamma_{\beta\alpha}$ and $\gamma_{\beta\beta}$ corresponding to the interference between PT-PR, PT-SR, ST-PR, and ST-SR respectively as also shown in Fig. \ref{fig:model}.

\subsection{Assumptions} 
We will assume that primary transmissions occur at a constant power and frequency for all PTs such that $\PP_i^\alpha = \PP^\alpha$ and $p_i^\alpha = p^\alpha \,\, \forall \, i$.  This will enable secondary transmitters to adapt their transmit power $\PP_i^\beta$ and their channel access probability $p_i^\beta$ given some local and global information about the primary network (including QoS requirements).
Moreover, in order to avoid any communication overheads and for the algorithm to be fully distributed we make the following simplification: 
all secondary transmitters assume that all other secondary transmitters will adapt their transmit powers and channel access probabilities in exactly the same way such that $\PP_i^\beta = \PP^\beta$ and $p_i^\beta = p^\beta \,\, \forall \, i$.

\subsection{Optimization Problem}
The goal is to maximize the number of secondary transmissions which achieve a target SINR of $q_\beta$.
The constraint is that primary pairs separated by a distance of no more than $R/2$ can achieve a target SINR of $q_\alpha$ with probability at least $Q$.
To this end, we first derive the outage probabilities for the primary and secondary network.
We then formulate the above distributed optimization problem and solve it numerically. Simulation results are later presented to demonstrate the performance of the algorithm.

\section{Outage Probability \label{sec:outage}}

\subsection{Derivation of the Connection Probability}

The outage probability $P_{out}$ is a fundamental performance metric of wireless networks and has been extensively studied in both cellular and mesh topologies.
For a given primary pair the connection probability $H_{ii}^{\alpha}=1-P_{out}$ is given by 
\es{
H_{i}^{\alpha}(d_{ii}^{\alpha\alpha}) &=\mathbb{P}[\mathrm{SINR}_{i}^{\alpha} \geq q_\alpha] \\
&= \mathbb{P}\Big[ |h_{ii}^{\alpha\alpha}|^2 \geq \frac{q_\alpha (\mathcal{N}+ \mathcal{I}_i^{\alpha\alpha} + \mathcal{I}_i^{\beta\alpha} )}{\mathcal{P}^\alpha g(d_{ij}^{\alpha\alpha})} \Big]
\label{H}}
and can be thought of as the probability that at any given instance of time, the link between the $i$th primary transmitter-receiver pair can achieve a target $\mathrm{SINR}$ of $q_\alpha$. 
Alternatively, equation \eqref{H} is the fraction of successfully received transmissions from node $i$ to node $j$, averaged over a long period of time.
The connection probability for the $i$th secondary transmitter-receiver pair $H_{i}^{\beta}$ is obtained by simply exchanging $\alpha \leftrightarrow \beta$ in \eqref{H}.

Conditioning on the interference realization $\mathcal{I}_i^{\alpha\alpha}$ and $\mathcal{I}_i^{\beta\alpha}$ and using the fact that $|h_{ii}^{\alpha\alpha}|^2 \sim \exp(1)$, the connection probability can be expressed as 
\es{
H_{i}^{\alpha} \! &= \! \mathbb{E}_{\mathcal{I}_i^{\alpha\alpha},\mathcal{I}_i^{\beta\alpha}} \Big[ \mathbb{P} \Big[ |h_{ii}^{\alpha\alpha}|^2 \! \geq \! \frac{q_\alpha (\mathcal{N}+ \!  \mathcal{I}_i^{\alpha\alpha} + \! \mathcal{I}_i^{\beta\alpha})}{\mathcal{P}^\alpha g(d_{ii}^{\alpha\alpha})} \Big| \mathcal{I}_i^{\alpha\alpha} , \mathcal{I}_i^{\beta\alpha}  \Big] \Big] \\
&= \mathbb{E}_{\mathcal{I}_i^{\alpha\alpha},\mathcal{I}_i^{\beta\alpha}} \prr{ \exp\pr{ -\frac{q_\alpha (\mathcal{N}+ \mathcal{I}_i^{\alpha\alpha} + \mathcal{I}_i^{\beta\alpha} )}{\mathcal{P}^{\alpha} g(d_{ii}^{\alpha\alpha})} } }\\
&= \exp\pr{-\frac{q_\alpha \mathcal{N}}{\mathcal{P}^\alpha g(d_{ii}^{\alpha\alpha})}} \mathcal{L}_{\mathcal{I}_i^{\alpha\alpha}} \pr{ s_\alpha /\mathcal{P}^\alpha} \mathcal{L}_{\mathcal{I}_i^{\beta\alpha}} \pr{ s_\alpha/\mathcal{P}^\alpha} 
\label{laplace}
,}
where $s_\alpha =\frac{q_\alpha }{ g(d_{ii}^{\alpha\alpha})}$ and
$\mathcal{L}_{\mathcal{I}_i^{xy}}(s_y/\mathcal{P}^\alpha) = \mathbb{E}_{\mathcal{I}_i^{xy}}\prr{e^{-s_y \mathcal{I}_i^{xy} /\mathcal{P}^\alpha} }$ is the Laplace transform of the random variable $\mathcal{I}_i^{xy}$ evaluated at $s_y/\mathcal{P}^\alpha$ conditioned on distance $d_{ii}^{yy}$.

The probability generating function for a Poisson point process $\Xi$ in $\R^2$ with intensity function $\lambda (\xi)$ \cite{streit2010poisson} satisfies 
\es{
\mathbb{E} \Big[ \prod_{\xi \in \Xi} f(\xi) \Big] = \exp\pr{ -\int_{\R^2} (1-f(\xi) ) \lambda(\xi) \dd \xi  }
\label{PGF2},}
for functions $f$ such that $0<f(\xi)\leq 1$.
We may therefore use \eqref{PGF2} to calculate the Laplace transforms in \eqref{laplace} to arrive at
\es{
\mathcal{L}_{\mathcal{I}_i^{\alpha\alpha}} \pr{ \frac{s_\alpha}{\mathcal{P}^\alpha}} &= \mathbb{E}_{|h_{ki}^{\alpha\alpha}|^2,d_{ki}^{\alpha\alpha},\chi_k^\alpha}\prr{e^{-s_\alpha \gamma_{\alpha\alpha}  \sum_{k\not=i} \chi_k^\alpha |h_{ki}^{\alpha\alpha}|^2 g(d_{ki}^{\alpha\alpha}) } }  \\ 
&= \exp\pr{- p^\alpha \lambda_\alpha \frac{2\pi^2 s_\alpha \gamma_{\alpha\alpha} (\epsilon +s_\alpha\gamma_{\alpha\alpha})^{\frac{2}{\eta}-1}}{\eta \sin \frac{2\pi}{\eta}}} \\
\mathcal{L}_{\mathcal{I}_i^{\beta\alpha}} \pr{ \frac{s_\alpha}{\mathcal{P}^\alpha}} &= \mathbb{E}_{|h_{ki}^{\beta\alpha}|^2,d_{ki}^{\beta\alpha},\chi_k^\beta}\prr{e^{- \frac{s_\alpha\gamma_{\beta\alpha}\PP^\beta}{\PP^\alpha} \sum_{k} \chi_k^\beta |h_{ki}^{\beta\alpha}|^2 g(d_{ki}^{\beta\alpha}) } }  \\ 
&= \exp\pr{\!- p^\beta \lambda_\beta \frac{2\pi^2 \frac{s_\alpha\gamma_{\beta\alpha}\PP^\beta}{\PP^\alpha} (\epsilon \!+ \!\frac{s_\alpha\gamma_{\beta\alpha}\PP^\beta}{\PP^\alpha})^{\frac{2}{\eta}-1}}{\eta \sin \frac{2\pi}{\eta}}}
\label{i}
}
where we have used the fact that the channel gains $|h_{ij}^{xy}|^2$ are independent and that the random access imposed by $\chi_k^x$ is equivalent to a Poisson thinning of the spatial intensity of transmitters by a factor of $p^x$.
The Laplace transforms $\mathcal{L}_{\mathcal{I}_i^{\beta\beta}} \pr{ s_\beta /\mathcal{P}^\beta}$ and $\mathcal{L}_{\mathcal{I}_i^{\alpha\beta}} \pr{ s_\beta /\mathcal{P}^\beta}$ needed to calculate $H_{i}^{\beta}$
can be obtained by simply exchanging $\alpha \leftrightarrow \beta$ in \eqref{i} and defining $s_\beta=\frac{q_\beta}{g(d_{ii}^{\beta\beta})}$.
These are standard results from stochastic geometry \cite{haenggi2009stochastic} in the case of $\epsilon=0$, generalized here for the non-singular path loss function \eqref{path} with $\epsilon>0$.

\textit{Remark 1: Notice that $H_i^\alpha$ and $H_i^\beta$ are exponentially decreasing functions of $p^\alpha$ and $p^\beta$.
That is, more secondary transmissions will increase the outage probability of both primary and secondary networks.
The same holds for primary transmissions.
Therefore, $p^\alpha H_i^\alpha$ and $p^\beta H_i^\beta$ have a unique maximum as functions of of $p^\alpha$ and $p^\beta$ respectively.}

\subsection{Numerical Verification of the Connection Probability}

\begin{figure}[t]
\centering
\includegraphics[width=0.45\textwidth]{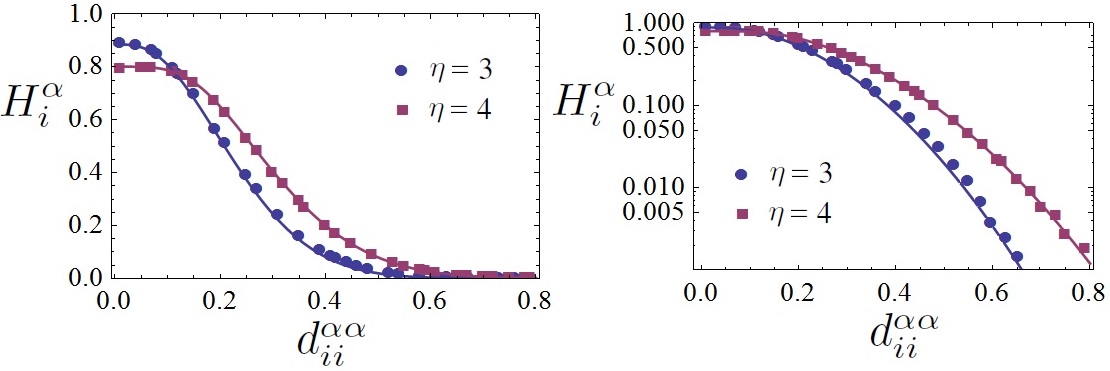}
\caption{
\textit{Left:} Comparison of computer simulations (point markers) with analytical equations \eqref{H} and \eqref{i} (smooth curves) for the connection probability $H_i^\alpha$ of primary transmitter-receiver pair separated by a distance $d_{ii}^{\alpha\alpha}\in[0,0.8]$. 
\textit{Right:} Same as left panel but on a log-linear scale.
Parameters used: $\PP^x = p^x = \lambda_x = \gamma_{xy} = q_x = \mathcal{N} = 1$ and $\epsilon=10^{-3}$ for $x,y=\{\alpha,\beta\}$.
}
\label{fig:num1}
\end{figure}

In this subsection we confirm equations \eqref{H} and \eqref{i} through computer simulations.
We briefly explain our simulation setup.
A large square region of side-length $L$ is defined with the center of the square as the coordinate origin.
Points are deployed in this region according to two independent Poisson point processes with intensities $\lambda_\alpha$ and $\lambda_\beta$ respectively such that the primary and secondary transmitter coordinates $\bt_k^\alpha$ and $\bt_k^\beta$ are uniformly distributed inside the square region.
Channel gains are chosen independently from an exponential distribution of mean $1$ and the interference $\mathcal{I}_i^{\alpha\alpha}$ and $\mathcal{I}_i^{\beta\alpha}$ are calculated at a receiver at the origin (i.e., $d_{ki}^{\alpha\alpha} = |\bt_k^\alpha |$ and $d_{ki}^{\beta\alpha} = |\bt_k^\beta |$).
Another random exponential variable is then chosen for $|h_{ii}^{\alpha\alpha}|^2$ and compared to $\frac{q_\alpha (\mathcal{N}+ \mathcal{I}_i^{\alpha\alpha} + \mathcal{I}_i^{\beta\alpha} )}{\mathcal{P}^\alpha g(d_{ij}^{\alpha\alpha})}$.
The above is then repeated $10^4$ times in a Monte Carlo loop in order to improve statistics on $H_i^\alpha$.
A second loop encompasses the above for different primary distances $d_{ii}^{\alpha\alpha}\in[0,0.8]$.
The simulation results are in perfect agreement with theory (equations \eqref{H} and \eqref{i}) and are shown in Fig. \ref{fig:num1} for different values of $\eta$.
Similar curves are observed for $H_i^\beta$ but are not shown here due to length restrictions.


\section{Distributed Optimization Scheme for Cognitive Radio\label{sec:CR}}

\begin{figure}[t]
\centering
\includegraphics[width=0.45\textwidth]{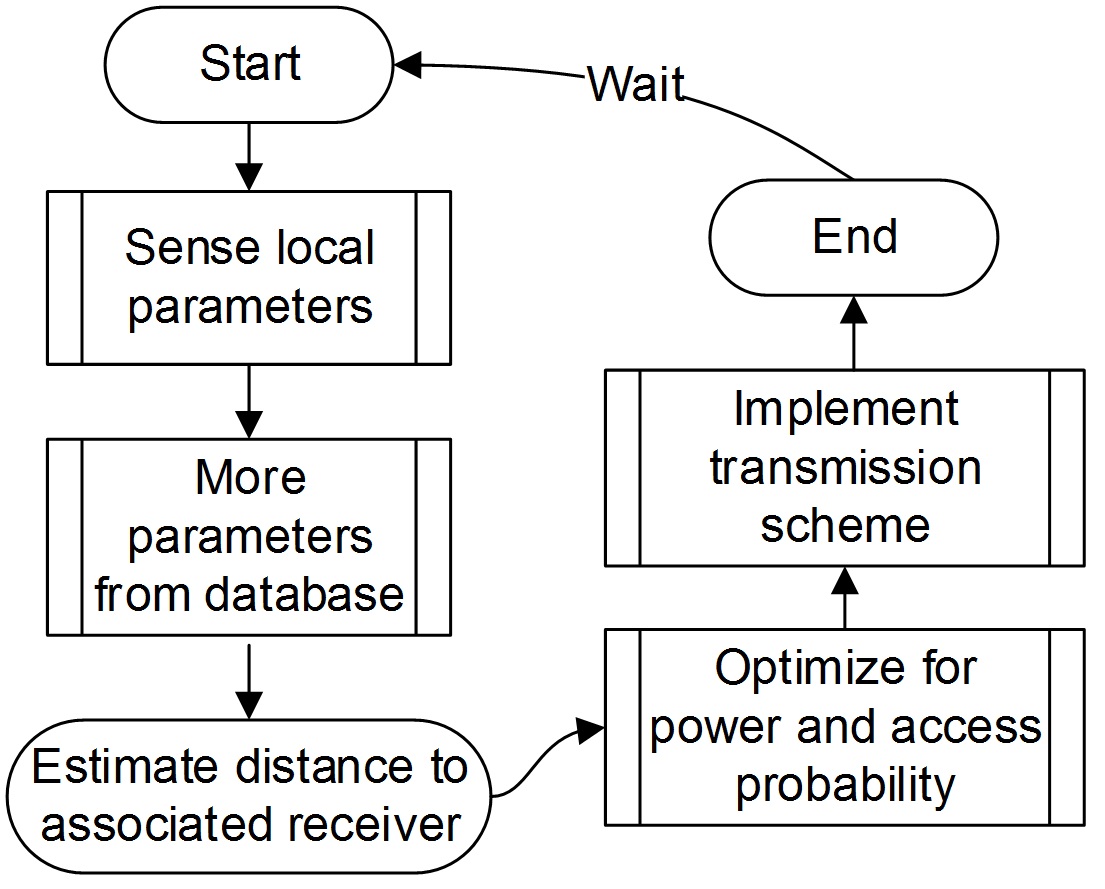}
\caption{Schematic of a simple process which can run the distributed power allocation and channel access probability assignment algorithm on each secondary transmitter.
}
\label{fig:method}
\end{figure}
\begin{figure}[ht]
\centering
\includegraphics[scale=0.3]{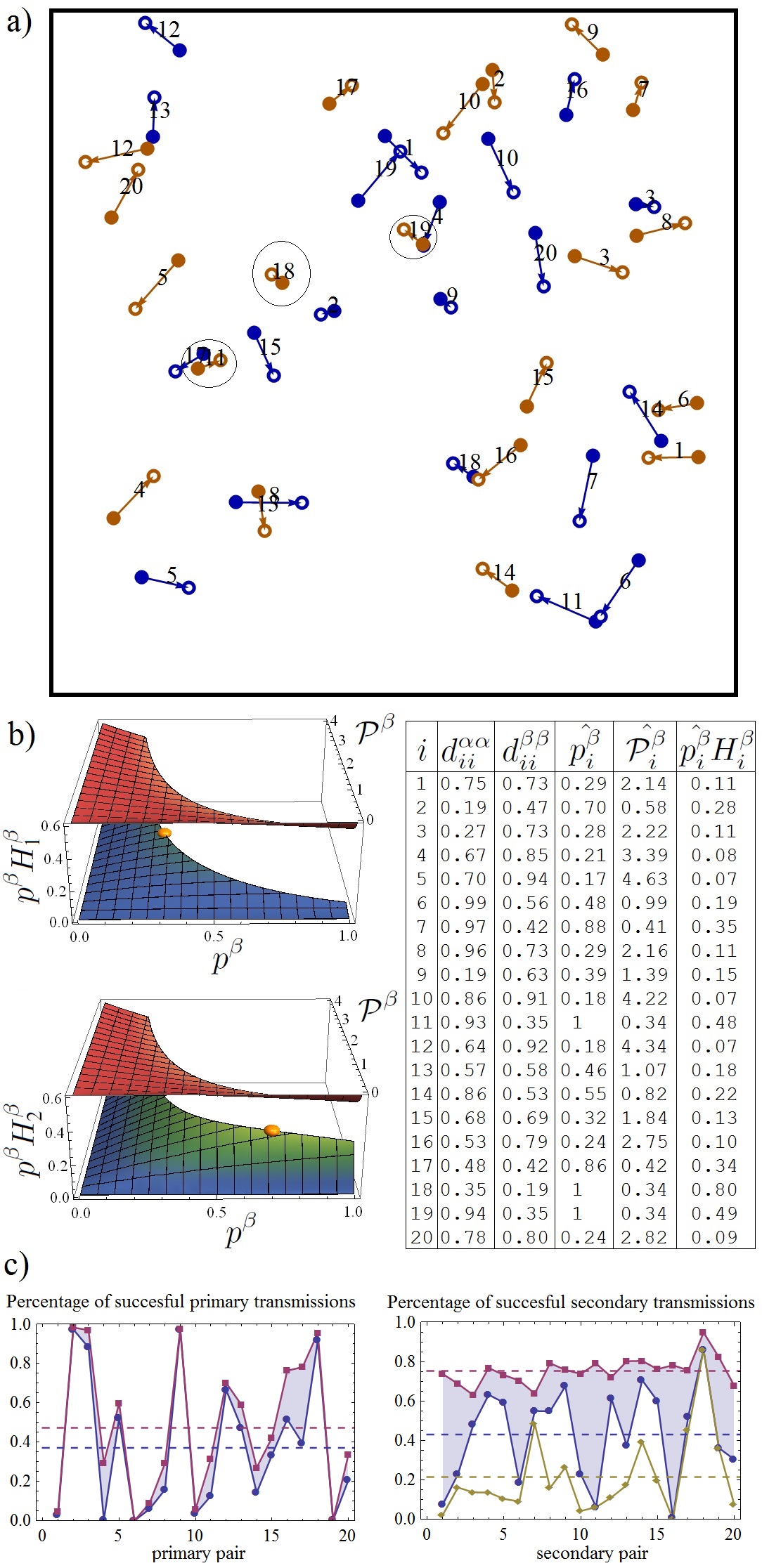}
\caption{
a) A schematic illustration of the setup described in Sec. \ref{setup}.
b) \textit{Left:} 3D plots of the probability of a successful secondary transmission $p^\beta H_i^\beta$ by users $i=1,2,$ (blue-green surface), and the probability of a successful primary transmission $H_i^\alpha (d_{ii}^{\alpha\alpha}=R/2)$ (red-orange surface) as a function of $p^\beta$ and $\PP^\beta$ using parameters as given in Sec. \ref{setup}.. 
Only the solution set satisfying $H_i^\alpha (d_{ii}^{\alpha\alpha}=R/2) > 50\%$ is plotted.
\textit{Right:} Table showing the distances between primary and secondary pairs, the optimal solutions $(\hat{\PP_i^\beta}, \hat{p_i^\beta})$ of \eqref{opt}. 
c) \textit{Left:} The percentage of successful primary and secondary transmissions for each node pair for the optimized secondary transmission strategy (purple square markers). 
Assuming no cross-tier interference this percentage is lower and is shown in blue circle markers.
\textit{Right:} Same as for the left, but for the secondary network pairs.
The blue markers are further normalized by the number of secondary transmissions.
The un-normalized results are shown using yellow diamonds markers.
}
\label{fig:num2}
\end{figure}

According to the above derivations, for the secondary transmitter $i$ to calculate its expected connection probability with its associated receiver $H_i^\beta$ it needs to know: 1) $\eta$ the path loss exponent characteristic of the local propagation medium, 2) $\PP^\alpha$ and $p^\alpha$ the PTs power and frequency of transmission respectively, 3) $\gamma_{\beta\beta}$ and $\gamma_{\alpha\beta}$ the code orthogonality between ST-SR and PT-SR pairs respectively, and 4) $d_{ii}^{\beta\beta}$ the distance to its associated SR. 
We will assume that STs have knowledge of these six parameters and formulate the optimization problem to be solved by each ST in a distributed fashion.

For a given secondary pair $i$, the number of transmissions which achieve a target SINR of $q_\beta$ over a long period of time is proportional to $p^\beta H_i^\beta$.
Moreover, in any given subset $\V\subset \R^2$ of area $V=|\V|$ and any given time instance one would expect to find $p^\beta \lambda_\beta  V$ active secondary transmitters.
Therefore, what each ST seeks to maximize is $p^\beta \lambda_\beta H_i^\beta$ which we term as the spatial density of successful secondary transmissions
\es{
(\hat{\PP_i^\beta}, \hat{p_i^\beta})= \argmax_{(\PP^\beta, p^\beta)} \prr{p^\beta \lambda_\beta H_i^\beta }
\label{opt}
,}
subject to the following three constraints 
\es{
H_{i}^\alpha \Big( d_{ii}^{\alpha\alpha}= \frac{R}{2} \Big) \geq Q , \qquad
\hat{\PP_i^\beta} \in[ \PP^\beta_{-} , \PP^\beta_{+}] , \qquad
\hat{ p_i^\beta} \in [0,1].
\label{const}}
Note that the first constraint in \eqref{const} is generic in that it does not depend on a specific primary pair. 
The second constraint is there to accommodate for any hardware restrictions of STs, whilst the third is trivial.
Similar metrics to the spatial density of successful secondary transmissions have been considered in the literature, including \cite{baccelli2006aloha,zhai2014local}.

\textit{Remark 2: Note that the solution $(\hat{\PP_i^\beta}, \hat{p_i^\beta})$ is unique to secondary transmitter $i$ and is different from that of $j\not= i$. 
The reason for this is that the solution for pair $i$ will depend on the distance $d_{ii}^{\beta\beta}$  whilst that of pair $j\not= i$ will depend on the distance $d_{jj}^{\beta\beta}$.
Nevertheless, since secondary receivers are identically and independently distributed it follows that the same holds for the separating distances between their associated transmitters.
Therefore, each secondary transmitter solves the optimization problem \eqref{opt} assuming that all other secondary users are separated by the same distance as himself; a reasonable guess in the absence of any communication or cooperation.
}

\section{Numerical Simulations and Performance Evaluation \label{sec:sim}}

\subsection{Simulation Setup}
In order to test the performance of the proposed distributed optimization approach we consider a simple two-tier wireless cognitive ad hoc network composed of hardware-constraint sensors.
Let the primary sensor network operate in some region of space tasked with collecting some environmental data e.g. temperature, pressure, humidity, pollution levels etc.
Moreover, assume that the primary sensor nodes are randomly deployed in this region of space.
The critical data recorded by each primary device is stored locally, and also transmitted to one nearby primary receiver for backup or redundancy purposes. 
This data is then collected periodically by a mobile master device whenever it is in range. 
This is the standard setup discussed in most WSN applications for agricultural, military, or industrial/building health monitoring purposes. 

Let us now assume that a secondary sensor network is to be deployed in the same geographical region and is to operate under the primary network and will also transmit its critical data to one nearby secondary receiver device.
Each secondary transmitting device must therefore optimize its transmission strategy (power and channel access) subject to some constraints without any cooperation from the primary network or other secondary users.
Let the optimization problem and constraints be exactly as defined in \eqref{opt} and \eqref{const} respectively.
The parameters required to solve this problem are as described in Sec. \ref{sec:CR} some of which can be sensed (e.g. $\eta, \gamma_{\beta\beta}, \gamma_{\alpha\beta}$ and $d_{ii}^{\beta\beta}$) and others which can be from some locally stored database (e.g. $p^\alpha$ and $\PP^\alpha$).
Each secondary transmitter can therefore run locally the simple process as depicted in Fig. \ref{fig:method}.

\subsection{Performance of numerical example \label{setup}}

To validate our method we now numerically simulate a network of $20$ primary and $20$ secondary node pairs (i.e., a total of $80$ nodes). 
We position the $40$ transmitting nodes randomly in a square domain ($10 \times 10$) such that $\lambda_\alpha = \lambda_\beta = 0.2$. 
The intended receiver nodes are randomly located within a radius of $R=1$ from their associated transmitters. 
A schematic of such a setup is shown in Fig. \ref{fig:num2}a).
Note that we keep length scales unit-less for simplicity since everything can be re-scaled according to a reference scale e.g the domain size.

We set $\gamma_{\beta\alpha}= \gamma_{\beta\beta} = \gamma_{\alpha\beta} = \gamma_{\alpha\alpha}= \PP^\alpha =p^\alpha = \mathcal{N} = q_\alpha= q_\beta=1$, $\PP^\beta_-=0$, $\PP^\beta_+=5$ and provide each secondary transmitter with the distance to its associated receiver such that it may numerically find the solution to \eqref{opt} and \eqref{const} for a given value of $Q \in [0,1]$.
This is a convex optimization problem and can be performed using standard gradient descent methods.
The solution set and the unique solution $(\hat{\PP_i^\beta}, \hat{p_i^\beta})$ for $Q= 50\%$ for secondary transmitters $\bt_1^\beta$ and $\bt_2^\beta$ are shown in Fig. \ref{fig:num2}b).
Moreover, the distances, the solutions, and the optimized variable are shown as a table in Fig. \ref{fig:num2}.
Notice that secondary pairs $i=11,18,$ and $19$ decide to transmit in every resource block ($\hat{p}_i^\beta =1$) but at a low power $\hat{\PP}_{i}^\beta \approx 0.34$.
We highlight these three in Fig. \ref{fig:num2}a).
It is clear that the reason for this is that $d_{ii}^{\beta\beta}$ is small.
Importantly however, only pair $18$ is sufficiently isolated from other nearby transmitters and therefore, we expect it to suffer less due to any interference.

Fig. \ref{fig:num2}c) shows the percentage of successful primary and secondary transmissions for each user pair.
Using the distances and secondary transmission strategies from the table in Fig. \ref{fig:num2}b) we simulate $10^5$ resource blocks, each time with a different realization of the Rayleigh fading channel (i.e., $|h_{ij}^{xy}|^2 \sim \exp(1)$) and count the number of times the primary and secondary pair $i=1,2\ldots 20$ has a successful transmission (i.e., achieves a target SINR of $q_\alpha$ and $q_\beta$ respectively).
For the secondary pairs these are normalized by the number of active transmissions of each transmitter (i.e., approximately $\hat{p}_i^\beta \times 10^5$ transmissions).
The results are shown using blue circle markers in Fig. \ref{fig:num2}c).
The purple square markers shows what would have been the case if $\gamma_{\alpha\beta}=\gamma_{\beta\alpha} =0$, i.e., if there was no cross-tier interference but using the same transmission strategy $(\hat{\PP_i^\beta}, \hat{p_i^\beta})$ as that given in the table of Fig. \ref{fig:num2}b).
Of course these are always higher than the blue circles as expected.
For the secondary pairs, the non-normalized results are shown using yellow diamonds.
Notice that secondary pairs $i=11,$ and $19$ are significantly affected by interference, whilst pair $18$ is not.
Looking at Fig. \ref{fig:num2}a) it is also clear that $\br_{16}^\beta \approx \bt_{18}^\alpha$ which is why the unlucky secondary pair $16$ has almost no successful transmissions.
Moreover, primary pair $19$ is significantly affected by primary transmitter $1$ and \textit{vice versa}.

The horizontal dashed lines in Fig. \ref{fig:num2}c) are the averages of the above performance metrics across all pairs.
These indicate that the primary network's performance has been affected by the interference caused by the secondary network by about $-27\%$.
An overall good outcome given that the secondary network operates at an average of $21.5\%$ successful transmissions per resource block per secondary transmitter, without any form of inter or intra tier cooperation and coordination.

\section{Discussion and Conclusions
\label{sec:con}}
This paper addressed the issue of distributed resource allocation and planning in non-cooperative cognitive radio networks.  The network connectivity, which is the complement of the outage probability, was derived using tools from stochastic geometry and was used in defining a network and a per-link performance metric.  
A simple process was introduced which can run on secondary (unlicensed) wireless transmitters with the aim to maximize the number of secondary transmissions per unit area, while not degrading primary (licensed) user average per-link performance.
The optimizations variables of interest are secondary user transmit power and channel access probability using locally sensed information as well as a look up table with primary network parameters.
The efficacy of the proposed approach and the validity of the derivations were demonstrated through numerical simulations.

\section*{Acknowledgements}
The authors would like to thank the directors of the Toshiba Telecommunications Research Laboratory for their support.


\bibliographystyle{ieeetr}
\bibliography{mybib}

\begin{thebibliography}{10}

\bibitem{Mitola1999a}
J.~Mitola and G.~Q. Maguire, ``Cognitive {R}adio: {M}aking {S}oftware {R}adios
  {M}ore {P}ersonal,'' {\em IEEE Personal Communications}, vol.~6, no.~4,
  pp.~13--18, 1999.

\bibitem{Cumanan2010}
K.~Cumanan, R.~Krishna, L.~Musavian, and S.~Lambotharan, ``{Joint Beamforming
  and User Maximization Techniques for Cognitive Radio Networks Based on Branch
  and Bound Method},'' no.~99, pp.~1--11, 2010.
\newblock Early Access.

\bibitem{DelRe2009}
E.~Del~Re, G.~Gorni, L.~Ronga, and R.~Suffritti, ``A {P}ower {A}llocation
  {S}trategy {U}sing {Game Theory in Cognitive Radio Networks},'' in {\em Proc.
  International Conference on Game Theory for Networks GameNets '09},
  pp.~117--123, 2009.

\bibitem{Fu2008}
F.~Fu and M.~van~der Schaar, ``{Stochastic {G}ame {F}ormulation for {C}ognitive
  {R}adio {N}etworks},'' in {\em Proc. 3rd IEEE Symposium on New Frontiers in
  Dynamic Spectrum Access Networks DySPAN 2008}, pp.~1--5, 2008.

\bibitem{haenggi2009stochastic}
M.~Haenggi, J.~Andrews, F.~Baccelli, O.~Dousse, and M.~Franceschetti,
  ``Stochastic geometry and random graphs for the analysis and design of
  wireless networks,'' {\em Selected Areas in Communications, IEEE Journal on},
  vol.~27, pp.~1029--1046, 2009.

\bibitem{Wu2008}
Y.~Wu and D.~Tsang, ``Distributed {M}ultichannel {P}ower {A}llocation
  {A}lgorithm for {S}pectrum {S}haring {C}ognitive {R}adio {N}etworks,'' in
  {\em Wireless Communications and Networking Conference, 2008. WCNC 2008.
  IEEE}, pp.~1436--1441, March 2008.

\bibitem{Yan2012}
F.~Yan, Y.~Yu, W.~Wang, and Y.~Zhang, ``Distributed {C}hannel and {P}ower
  {A}llocation {B}ased on {H}ybrid {S}pectrum {A}ccess in {C}ognitive {R}adio
  {N}etwork,'' in {\em Cloud Computing and Intelligent Systems (CCIS), 2012
  IEEE 2nd International Conference on}, vol.~02, pp.~898--903, Oct 2012.

\bibitem{Wu2009}
Y.~Wu and D.~Tsang, ``Distributed {P}ower {A}llocation {A}lgorithm for
  {S}pectrum {S}haring {C}ognitive {R}adio {N}etworks with {QoS} {G}uarantee,''
  in {\em INFOCOM 2009, IEEE}, pp.~981--989, April 2009.

\bibitem{Seneviratne2011}
C.~Seneviratne and H.~Leung, ``A {G}ame {T}heoretic {A}pproach for {R}esource
  {A}llocation in {C}ognitive {W}ireless {S}ensor {N}etworks,'' in {\em
  Systems, Man, and Cybernetics (SMC), 2011 IEEE International Conference on},
  pp.~1992--1997, Oct 2011.

\bibitem{Ahmed2014}
F.~Ahmed, O.~Tirkkonen, A.~Dowhuszko, and M.~Juntti, ``Distributed {P}ower
  {A}llocation in {C}ognitive {R}adio {N}etworks under {N}etwork {P}ower
  {C}onstraint,'' in {\em Cognitive Radio Oriented Wireless Networks and
  Communications (CROWNCOM), 2014 9th International Conference on},
  pp.~492--497, June 2014.

\bibitem{Nadkar2012b}
T.~Nadkar, V.~Thumar, G.~Tej, S.~Merchant, and U.~Desai, ``Distributed {P}ower
  {A}llocation for {S}econdary {U}sers in a {C}ognitive {R}adio {S}cenario,''
  {\em Wireless Communications, IEEE Transactions on}, vol.~11, pp.~1576--1586,
  April 2012.

\bibitem{georgiou2014network}
O.~Georgiou, C.~P. Dettmann, and J.~P. Coon, ``Network connectivity: Stochastic
  vs. deterministic wireless channels,'' in {\em Communications (ICC), 2014
  IEEE International Conference on}, pp.~77--82, IEEE, 2014.

\bibitem{keyhole}
O.~Georgiou, M.~Bocus, M.~Rahman, C.~Dettmann, and J.~Coon, ``Network
  connectivity in non-convex domains with reflections,'' {\em Communications
  Letters, IEEE}, vol.~19, pp.~427--430, March 2015.

\bibitem{dousse2004connectivity}
O.~Dousse and P.~Thiran, ``Connectivity vs capacity in dense ad hoc networks,''
  in {\em INFOCOM 2004. Twenty-third AnnualJoint Conference of the IEEE
  Computer and Communications Societies}, vol.~1, IEEE, 2004.

\bibitem{gong2014interference}
Z.~Gong and M.~Haenggi, ``Interference and outage in mobile random networks:
  Expectation, distribution, and correlation,'' {\em Mobile Computing, IEEE
  Transactions on}, vol.~13, no.~2, pp.~337--349, 2014.

\bibitem{metcalfe1976ethernet}
R.~M. Metcalfe and D.~R. Boggs, ``Ethernet: distributed packet switching for
  local computer networks,'' {\em Communications of the ACM}, vol.~19, no.~7,
  pp.~395--404, 1976.

\bibitem{kleinrock1975packet}
L.~Kleinrock and F.~A. Tobagi, ``Packet switching in radio channels: Part
  i--carrier sense multiple-access modes and their throughput-delay
  characteristics,'' {\em Communications, IEEE Transactions on}, vol.~23,
  no.~12, pp.~1400--1416, 1975.

\bibitem{baccelli2006aloha}
F.~Baccelli, B.~Blaszczyszyn, and P.~Muhlethaler, ``An aloha protocol for
  multihop mobile wireless networks,'' {\em Information Theory, IEEE
  Transactions on}, vol.~52, no.~2, pp.~421--436, 2006.

\bibitem{yang2012connectivity}
T.~Yang, G.~Mao, and W.~Zhang, ``Connectivity of large-scale csma networks,''
  {\em Wireless Communications, IEEE Transactions on}, vol.~11, no.~6,
  pp.~2266--2275, 2012.

\bibitem{streit2010poisson}
R.~L. Streit, {\em Poisson Point Processes}, vol.~1.
\newblock Springer, 2010.

\bibitem{zhai2014local}
D.~Zhai, M.~Sheng, X.~Wang, and Y.~Zhang, ``Local connectivity of cognitive
  radio ad hoc networks,'' in {\em Global Communications Conference (GLOBECOM),
  2014 IEEE}, pp.~1078--1083, IEEE, 2014.

\end{thebibliography}

\end{document}